# Non-Markovian Exceptional Points by Interpolating Quantum Channels

Wai Chun Wong[1], Bei Zeng[2,*], and Jensen Li[1,*]

[1]*Department of Engineering, University of Exeter, EX4 4QF, UK*
[2] *Department of Physics, The University of Texas at Dallas, Richardson, Texas 75080, USA*

Exceptional points (EPs) are special points in non-Hermitian systems where both eigenvalues and eigenvectors coalesce. In open quantum systems, these points are typically analyzed using effective non-Hermitian Hamiltonians or Liouvillian superoperators. While quantum channels offer the most general framework for describing state evolution in such systems, the existence and properties of EPs within this setting remain largely unexplored. In this work, we present a general strategy for generating quantum EPs for a single-qubit setting. We show that quantum channels can be separated into two distinct phases, with the transition between them marked by the presence of an EP. Based on this, we propose a systematic method to realize EPs by interpolating between quantum channels representing different phases. Experimentally, we implement these interpolated channels on a nuclear magnetic resonance (NMR) quantum computer and confirm the emergence of second-order EPs with high fidelity. Extending the interpolation to three channels further reveals third-order EPs. Our results establish quantum channel interpolation as a versatile framework for generating EPs and provide a general description of EPs in open quantum systems.

* j.li13@exeter.ac.uk, Bei.Zeng@UTDallas.edu

Non-Hermitian physics, which describes systems with dissipation and amplification, has gained significant attention in recent years [1-5]. A key example is parity-time ($\mathcal{PT}$)-symmetric systems, where balanced gain and loss enable real eigenvalues despite non-Hermiticity, leading to exceptional points (EPs) where eigenvalues and eigenvectors coalesce [6, 7]. EPs have been widely studied in optics, acoustics, and electronic systems, enabling unidirectional invisibility, novel lasing behaviors, and enhanced sensing [8–20]. In passive systems, introducing a global loss bias allows PT-like behavior without requiring gain [8], while state evolution around EPs exhibits path-dependent transport effects [21-22].

Recently, EPs have been extended to quantum open systems, with demonstrations of chiral state transfer in trapped ions and cold atoms [23-25]. These implementations often rely on engineering an effective PT-symmetric Hamiltonian to obtain Hamiltonian exceptional points (HEPs), but this approach does not fully capture quantum dynamics in the presence of decoherence or noise [26-28]. A more complete description requires the Lindblad master equation, where EPs emerge in the non-Hermitian Liouvillian superoperator, termed Liouvillian exceptional points (LEPs). Unlike HEPs, LEPs can arise even when no corresponding HEPs exist, offering new opportunities for quantum state control [29-32]. Experimental realizations of LEPs have been reported in trapped ions, superconducting qubits, and quantum computers [27-34]. However, these studies generally assume Markovian environments to ensure a well-defined Lindblad master equation, whereas recent work has proposed EPs in non-Markovian regimes, where the Liouvillian may not always remain well defined [35].

Here, we propose interpolating quantum channels as a versatile framework for establishing EPs in open quantum systems [36]. These quantum channels are completely positive and trace-preserving (CPTP) in a single step but are generally non-divisible, indicating non-Markovian dynamics. Consequently, their Liouvillian does not necessarily remain positive at all time steps.

Notably, such a quantum channel is guaranteed to exist in one of two distinct phases—one characterized by purely real eigenvalues and the other by complex conjugate pairs—without requiring symmetry constraints such as PT symmetry. By interpolating between quantum channels from different phases, EPs naturally emerge at the transition, providing a systematic method for generating them. We apply this approach to a single-qubit channel to construct both second- and third-order EPs, corresponding to the interpolation of two and three channels, respectively. In the latter case, EP lines also emerge. Using quantum process tomography, previously employed to identify LEPs [34], we demonstrate non-Markovian EPs realized through our channel interpolation approach on a nuclear magnetic resonance (NMR) quantum computer.

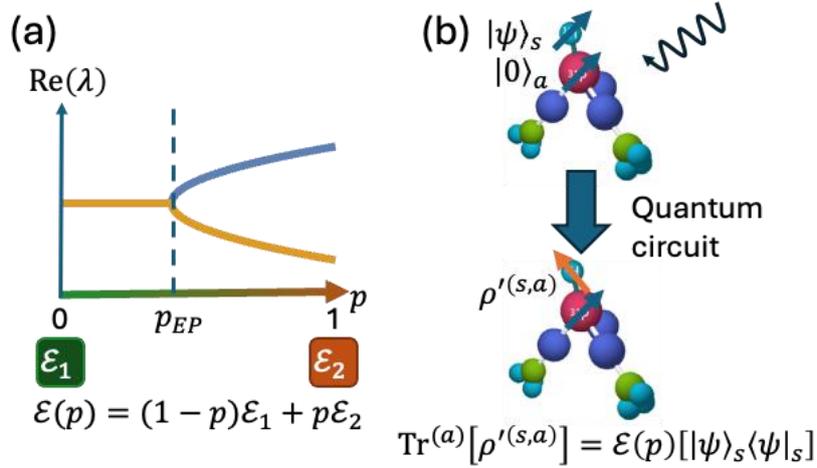

FIG. 1. (a) Schematic of linear interpolation between two quantum channels to generate a family of quantum channels with exceptional points. Not all eigenvalues are included for clarity. Two quantum channels $\mathcal{E}_1$ and $\mathcal{E}_2$ are connected by a linear interpolation $\mathcal{E}(p) = (1-p)\mathcal{E}_1 + p\mathcal{E}_2$ with $p \in [0,1]$. The exceptional point is located at $p = p_{EP}$ where the eigenvalues of the superoperator $\mathcal{E}(p)$ coalesce. (b) Simulation of the quantum channels with a circuit-based nuclear magnetic resonance (NMR) quantum computer. Two qubits are separated as signal and ancilla qubits. The signal qubit and ancilla qubit are initialized to the state $|\psi\rangle_s$ and $|0\rangle_a$ respectively. After the quantum circuit, the signal qubit is measured to obtain the output state of the quantum channel $\mathcal{E}(p)[|\psi\rangle_s\langle\psi|_s]$.

Open quantum systems interact with their environment, leading to decoherence and information loss beyond unitary dynamics. The Markovian approximation captures these effects through the Lindblad master equation $d\rho/dt = \mathcal{L}\rho$, where the Liouvillian superoperator $\mathcal{L}$ governs how quantum states evolve under environmental influence. For a single qubit, this evolution can be represented by a 4×4 non-Hermitian matrix acting on vectorized density matrix $\{\rho_{11}, \rho_{21}, \rho_{12}, \rho_{22}\}$. This non-Hermitian structure allows for the emergence of Liouvillian exceptional points (LEPs): parameter values where eigenvalues and eigenvectors coalesce, fundamentally altering the system's dynamics.

We extend this exceptional point physics beyond Markovian dynamics to general quantum channels. While Lindbladian evolution assumes memoryless (Markovian) environments, quantum channels $\mathcal{E}$ describe a broader range of quantum state transformations, including those with memory effects, non-Markovian noise, and measurement backaction. By definition, quantum channel $\mathcal{E}$ are maps satisfying complete positivity and trace preservation (CPTP):

$$\begin{pmatrix} \rho'_{11} \\ \rho'_{21} \\ \rho'_{12} \\ \rho'_{22} \end{pmatrix} = \mathcal{E} \begin{pmatrix} \rho_{11} \\ \rho_{21} \\ \rho_{12} \\ \rho_{22} \end{pmatrix}. \qquad (1)$$

with the transformed density matrix $\rho' = \begin{pmatrix} \rho'_{11} & \rho'_{12} \\ \rho'_{21} & \rho'_{22} \end{pmatrix}$. For Markovian evolution over time $T$, the channel takes the form $\mathcal{E} = e^{\mathcal{L}T}$. This connection motivates us to introduce channel exceptional points as degeneracies in the eigenvalues and eigenvectors of the channel map $\mathcal{E}$ itself. This generalization is natural because every LEP automatically induces a channel exceptional point through the exponential map $\mathcal{E} = e^{\mathcal{L}T}$. However, the concept is more general, many quantum channels cannot be expressed as time evolution under any master equation, yet they can still exhibit exceptional points.

This extension opens new territory. While LEPs are confined to Markovian dynamics, channel exceptional points exist across all CPTP maps, including non-Markovian processes and measurement-induced evolution. By studying exceptional points at the channel level, we gain access to a broader landscape of eigenvalues degeneracies in quantum regime.

To revel two distinct phases of a quantum channel, we change the basis in Eq. (1) to

$$\begin{pmatrix} 1 \\ r'_x \\ r'_y \\ r'_z \end{pmatrix} = \begin{pmatrix} 1 & 0 & 0 & 0 \\ s_x & E_{xx} & E_{xy} & E_{xz} \\ s_y & E_{yx} & E_{yy} & E_{yz} \\ s_z & E_{zx} & E_{zy} & E_{zz} \end{pmatrix} \begin{pmatrix} 1 \\ r_x \\ r_y \\ r_z \end{pmatrix}. \qquad (2)$$

Here, the quantum channel is represented by transforming Bloch vector components $r_x = \rho_{12} + \rho_{21}$, $r_y = i(\rho_{21} - \rho_{12})$ and $r_z = \rho_{11} - \rho_{22}$. Since these components, along with $1 = \rho_{11} + \rho_{22}$ are linear in terms of $\{\rho_{11}, \rho_{21}, \rho_{12}, \rho_{22}\}$, the transformation matrix is given by $M\mathcal{E}M^{-1}$ with an invertible matrix and they share the same eigenvalues and channel EPs. For simplicity, we will also refer to this matrix as $\mathcal{E}$ when the context is clear. The first row, fixed as $(1,0,0,0)$ due to the trace-preserving condition, ensures a trivial eigenvalue of 1 and other eigenvalues given by eigenvalues of submatrix $E$. Geometrically this distortion matrix $E$ map the Bloch vector of state $\rho$ and shifted by the shift vector $s$. Since $\mathcal{E}$ is now a real matrix, its eigenvalues appear in conjugate pairs, as $\mathcal{E}v = \lambda v$ implies $\mathcal{E}^*v^* = \mathcal{E}v^* = \lambda^*v^*$ where $*$ represent complex conjugate. The eigenvalues can be either fully real, $\{1, \lambda_1, \lambda_2, \lambda_3\} \in \mathbb{R}$ or form one conjugate pair i.e. $\{1, \lambda_1\} \in \mathbb{R}, \{\lambda_2, \lambda_2^*\} \in \mathbb{C}$.

It is easily to show that in the first case, all eigenvectors are real, while in the second, the eigenvector of $\lambda_1$ is real, and those of $\{\lambda_2, \lambda_2^*\}$ form a conjugate pair if the eigenvalues are non-degenerate. The degenerate case will be considered later. This mirrors the behavior of $\mathcal{PT}$-symmetric systems, where eigenvalues are real in the exact phase and complex in the broken phase. However, in our case, these phases arise from the real nature of $\mathcal{E}$ ($\mathcal{E} = \mathcal{E}^*$) rather than an engineered gain-loss balance in $\mathcal{PT}$-symmetric systems. To distinguish this from $\mathcal{PT}$-symmetry, we introduce the complex conjugation operator $\mathcal{K}$, which emphasize the real nature of the channel rather than physical time reversal. This allows us to classify the system's phases as the $\mathcal{K}$-exact phase, where eigenvalues and eigenvectors are real, and the $\mathcal{K}$-broken phase, where eigenvalues and eigenvectors form complex conjugate pairs.

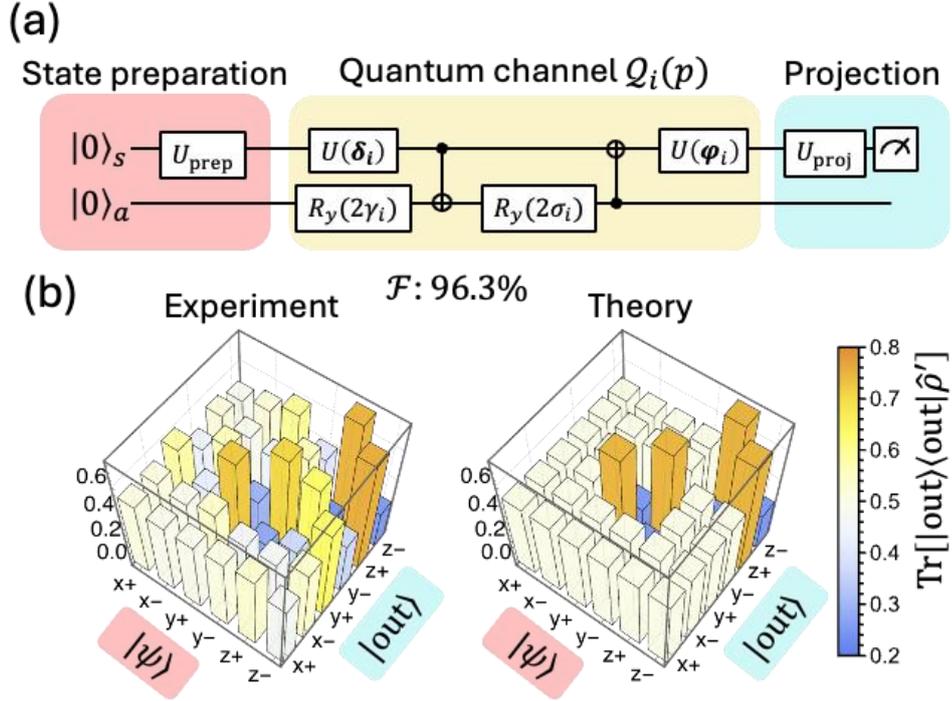

FIG. 2. (a) Circuit for the implementation of the quantum channels $\mathcal{E}(p)$ and process tomography. Each quantum channel $\mathcal{E}(p)$ is implemented by splitting into two simpler quantum channels $\mathcal{Q}_1(p), \mathcal{Q}_2(p)$ which can be implemented using 2 qubits with U3 gates, Ry gates and CNOT gates shown in the middle. The process tomography is performed by preparing the input state $|\psi_s\rangle$ into eigenstates of Pauli operators, i.e., $\{|x_\pm\rangle, |y_\pm\rangle, |z_\pm\rangle\}$ using $U_{\text{prep}}$ and measuring measure Pauli observables using $U_{\text{proj}}$. (b) Experimental results of the measurement for the quantum channels $\mathcal{E}(p=1)$ and the theory prediction. We obtain a fidelity of 96.3% for the experimental result.

To examine the phase transition, we consider a linear interpolation between two quantum channels, $\mathcal{E}_1$ and $\mathcal{E}_2$ with different phases, as shown in Fig. 1(a):

$$\mathcal{E}(t) = (1-p)\mathcal{E}_1 + p\mathcal{E}_2 \qquad (3)$$

where $p \in [0,1]$. Physically, this represents a state entering $\mathcal{E}_1$ with probability $1-p$ and $\mathcal{E}_2$ with probability $p$. As an example, we consider two quantum channels with vanishing shift vector $s = 0$, where their distortion matrix $E$ are given by

$$E_1 = \frac{1}{2}\begin{pmatrix} 0 & 0 & 0 \\ 0 & 0 & -1 \\ 0 & 1 & 0 \end{pmatrix}, \text{ and } E_2 = \frac{1}{2}\begin{pmatrix} 0 & 0 & 0 \\ 0 & 1 & 0 \\ 0 & 0 & -1 \end{pmatrix} \qquad (4)$$

These channels are chosen for their simplicity and their distinct phases. While their complete positivity (CP) is not immediately apparent from the matrix form, it becomes clear from their Kraus representations:

$$\mathcal{E}_1[\rho] = \frac{1}{2}\sqrt{\sigma_x}\rho\sqrt{\sigma_x}^\dagger + \frac{1}{4}\sigma_y\rho\sigma_y + \frac{1}{4}\sigma_z\rho\sigma_z$$
$$\mathcal{E}_2[\rho] = \frac{1}{4}\rho + \frac{1}{4}\sigma_x\rho\sigma_x + \frac{1}{2}\sigma_y\rho\sigma_y \quad (5)$$

Since each term is CP, their sum is CP as well (TP can be seen from Eq. (2)). To analyze the EPs in the interpolated quantum channel, we examine how the distortion matrix varies with $p$. By linearity, the distortion matrix of $\mathcal{E}(p)$ is

$$E(p) = (1-p)E_1 + pE_2 = \frac{1}{2}\begin{pmatrix} 0 & 0 & 0 \\ 0 & p & p-1 \\ 0 & 1-p & -p \end{pmatrix} \quad (6)$$

The eigenvalues are $\lambda_1 = 0$ and $\lambda_\pm = \pm\sqrt{p/2 - 1/4}$, with eigenvectors $v_1 = \{1,0,0\}$ and $v_\pm = \{0, p + 2\lambda_\pm, 1-p\}$). Both eigenvectors coalesce at $p = 1/2$, indicating an exceptional point.

Although we choose the quantum channels $\mathcal{E}_0$ and $\mathcal{E}_1$ for simplicity, the concept extends generally. Linear interpolation between any two quantum channels with different phases inevitably leads to phase transitions where eigenvalues coalesce, indicating potential EPs. At these transitions, two cases arise: true EPs, where both eigenvalues and eigenvectors coalesce, and Diabolic Points (DPs), where only eigenvalues merge while eigenvectors remain distinct. This insight suggests a practical approach for generating EPs by systematically testing pairs of quantum channels with different phases. If a transition results in a DP rather than an EP, one can simply try another pair. Given the vast number of quantum channels and the simplicity of this procedure, this provides an efficient method for generating EPs in quantum channels.

To experimentally simulate the exceptional point, we use a two-qubit nuclear magnetic resonance (NMR) circuit-based quantum computer to implement the quantum channels $\mathcal{E}(p)$ and retrieve their eigenvalues. The standard approach for implementing a single-qubit quantum channel, Stinespring dilation [37], requires two additional ancillary qubits (totally 3 needed) and the implementation of a general three-qubit unitary gate, which is challenging even for superconducting quantum computers in the context of simulating EPs [34]. Instead, we follow the method from [38] decomposing $\mathcal{E}(p)$ into two simpler quantum channels $Q_1(p), Q_2(p)$ such that $\mathcal{E}(p) = 1/2(Q_1(p) + Q_2(p))$ where each of these channels requires only one ancillary qubit (totally 2 needed) and can be implemented using the quantum circuit shown in Fig. 2(a). This circuit consists of two $U_3$ gates ($U(\delta), U(\varphi)$) for diagonalizing the distortion matrix, along with two $R_y$ and two CNOT gates to implement shift vector and diagonalized distortion matrix. We implement $Q_1(p)$ and $Q_2(p)$ separately, averaging their measurement results to construct $\mathcal{E}(p)$. The algorithm for determining the circuit parameters is detailed in Appendix A.

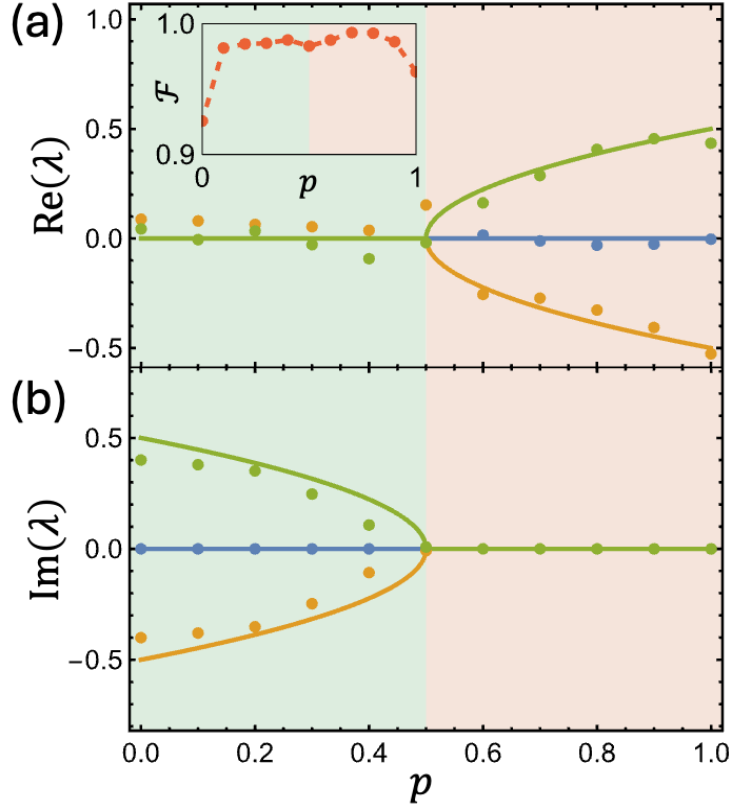

FIG. 3. (a) The real and (b) imaginary parts of the eigenvalues of the superoperator $\mathcal{E}(p)$ as a function of $p$. The trivial eigenvalue 1 is omitted for clarity. The solid lines are the theoretical prediction and the dots are the experimental results by the quantum process tomography. The exceptional point is located at $p = 0.5$ where the two eigenvalues coalesce. Inset: The fidelity of the experimental results with the theoretical prediction. The fidelity is above 93% for the whole range of $p$.

To obtain the eigenvalues, we perform quantum process tomography (QPT) to fully characterize the superoperator $\mathcal{E}(p)$. For each quantum channel $\mathcal{E}(p)$, we prepare six input states corresponding to the eigenstates of the Pauli operators, i.e. $\{|x_\pm\rangle = (|0\rangle \pm |1\rangle)/\sqrt{2}, |y_\pm\rangle = (|0\rangle \pm i|1\rangle)/\sqrt{2}, |z_+\rangle = |0\rangle, |z_-\rangle = |1\rangle\}$, achieved by applying appropriate state preparation unitary $U_{\text{prep}}$. We then measure the output states in these same Pauli bases by applying suitable measurement unitary $U_{\text{proj}}$ before readout. The Total circuit for the implementation of the quantum channels $Q_{1,2}(p)$ and the process tomography is shown in Fig. 2(a).

The experiment result for $\mathcal{E}(1) = \mathcal{E}_2$ is shown in Fig. 2(b) with high agreement with the theory prediction. For the channel tomography, we use this result to perform maximum-likelihood fitting for a quantum channel with CPTP condition [39]. We found that the process fidelity $\mathcal{F}$ of the experimental result with the theoretical prediction is 96.3% [40]. The fitting process and the definition of the fidelity of quantum channel is provided in appendix B. After reconstructing the quantum channel $\mathcal{E}(p)$, we plot its eigenvalues as a function of $p$ in Fig. 3. The experimental results are in good agreement with the theoretical prediction, and the exceptional point is located at $p = 0.5$ where the two eigenvalues coalesce. The fidelity of the experimental results with the theoretical prediction is above 93% for the whole range of $p$ as shown in the inset of Fig. 3. This demonstrates the existence of exceptional points in the quantum channels $\mathcal{E}(p)$ and the successful implementation of the quantum channels using a circuit-based NMR quantum computer.

The previous example focuses on interpolating 2 quantum channels. We can further extend our scheme to interpolating 3 quantum channels. With an extra channel we now have one more interpolating parameter where we expect richer phenomenon at the phase transition such as higher order exceptional point as we shall see. We consider 3 quantum channels $\mathcal{E}_1$, $\mathcal{E}_2$ and $\mathcal{E}_3$ and interpolating them by

$$\mathcal{E}(a_1, a_2, a_3) = a_1 \mathcal{E}_1 + a_2 \mathcal{E}_2 + a_3 \mathcal{E}_3 \tag{7}$$

with $a_1 + a_2 + a_3 = 1$. We choose the quantum channels $\mathcal{E}_1$ and $\mathcal{E}_2$ as same as before and for the third quantum channel, we chosen to have the distortion matrix $\mathcal{E}_3$ as a rotation matrix for the rotation along as an axis $\hat{n} = \{1,1,1\}$ for angle $-\pi/2$ to further mix all three principle axes. By Rodrigues' rotation formula we have the distortion matrix $\boldsymbol{E}_3$ as

$$\boldsymbol{E}_3 = \frac{1}{3}\begin{pmatrix} 1 & 1+\sqrt{3} & 1-\sqrt{3} \\ 1-\sqrt{3} & 1 & 1+\sqrt{3} \\ 1+\sqrt{3} & 1-\sqrt{3} & 1 \end{pmatrix} \tag{8}$$

In Fig. 4(a), we show the phase diagram of the 3 quantum channels interpolation. At the base of the triangle ($a_3 = 0$), we have the same linear interpolation as before and the phase transition is now a line in the parameter space. We have calculated the phase rigidity and confirmed that these phase transition lines are indeed exceptional points, as at least one phase rigidity vanishes. We found that there is convergence of 2 phase transition line at $(a_1: a_2: a_3) = (10: 2\sqrt{13}: 3\sqrt{3})$.

At this convergence point we found a distortion matrix having 3 coalescing eigenvalues and eigenvectors. This is an order 3 exceptional point EP$_3$. Here we do not perform the experiment for the 3 quantum channels interpolation, but the experimental implementation is feasible and can be done using the same method as the 2 quantum channels interpolation. In Fig. 4(b), we show the real and imaginary parts of the eigenvalues of the quantum channel $\mathcal{E}(a_1, a_2, a_3)$ as a function of $a_1$ and fixing $a_2$ around the exceptional point. We can observe that the 2 order 2 exceptional points coalesce and form an order 3 exceptional point.

We should note that our work can be extended to higher dimensional quantum channels. In fact, the Pauli basis used to describe the quantum channel have a natural extension to higher dimension. For example, in 2 qubits case, the Pauli basis consists of 16 operators including the identity operator and 15 tensor products of Pauli operators. The quantum channel can be represented as a $16 \times 16$ real matrix in this basis. Then the eigenvalues will be still either all real or some of them form complex conjugate pairs. However, unlike the single qubit case, the number of complex conjugate pairs can be more than 1. Perhaps we can classify the quantum channel into different phases by the number of conjugate pairs and the exceptional point will only occur at phase transition. This implies exceptional points in higher dimensional quantum channels can be more sophisticated and richer than the single qubit case and worth investigation in further studies. The importance of understanding higher dimensional quantum channels is highlighted in recent studies of quantum thermal machines [31], where two-qubit systems serve as a models for quantum heat engines In such systems, the interplay between exceptional points and the system dynamics could lead to novel ways of controlling the quantum thermal machines.

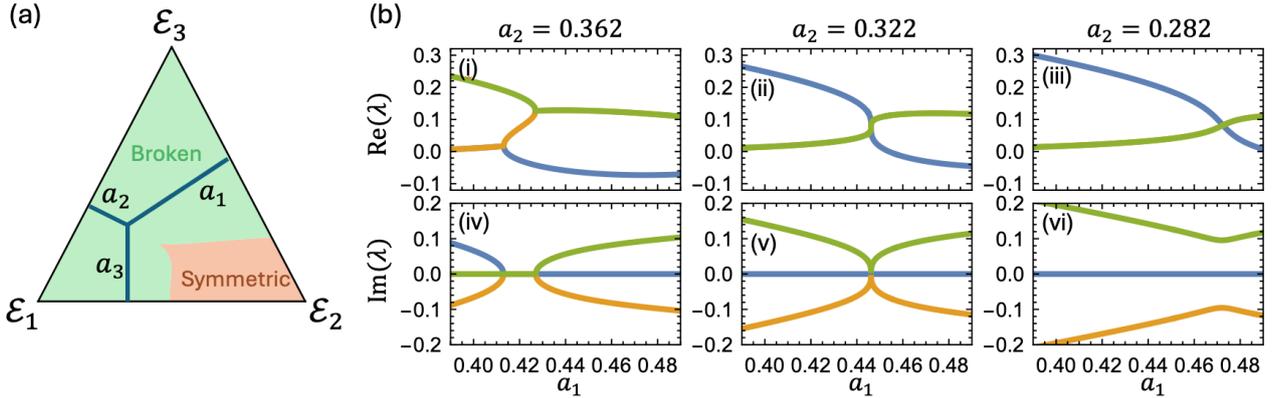

FIG. 4. (a) Phase diagram for interpolating 3 quantum channels $\mathcal{E}_1$, $\mathcal{E}_2$ and $\mathcal{E}_3$ with distortion matrix $E_1$, $E_2$, and $E_3$ in Eq. (4) and (8). The interpolating parameters are $a_1$, $a_2$ and $a_3$ under the constraint $a_1 + a_2 + a_3 = 1$ to ensure the resultant quantum channel $\mathcal{E}(a_1, a_2, a_3) = a_1\mathcal{E}_1 + a_2\mathcal{E}_2 + a_3\mathcal{E}_3$ is a valid quantum channel. The triangle represents the convex hull of these 3 quantum channels. There is an order 3 exceptional point EP$_3$ located at $(a_1, a_2, a_3) \simeq (0.446, 0.322, 0.232)$ (b) The real (i), (iii), (v) and imaginary parts (ii), (iv), (vi) of the eigenvalues of the superoperator $\mathcal{E}$ $(a_1, a_2, a_3)$ as a function of $a_1$ with fixed $a_2 = \{0.362, 0.322, 0.282\}$ and $a_3 = 1 - a_1 - a_2$. The trivial eigenvalue 1 is omitted for clarity. When $a_x$ decrease from 0.361 to 0.321, 2 order 2 exceptional point EP$_2$ coalesce and form an order 3 exceptional point EP$_3$ at $(a_1, a_2, a_3) \simeq (0.446, 0.322, 0.232)$.

In conclusion, we have demonstrated a systematic framework for generating and studying exceptional points in quantum channels through linear interpolation. Our experimental implementation on an NMR quantum computer achieved process fidelities above 96%, confirming the existence of second-order exceptional points at predicted parameter values. By extending the interpolation to three quantum channels, we revealed higher-order exceptional points, including a third-order EP at specific interpolation parameters. This approach establishes quantum channel interpolation as a versatile tool for generating and studying exceptional points in open quantum systems. Beyond demonstrating new sources of exceptional points, our work provides a general framework for understanding phase transitions in quantum channels. Future work could explore these channel-based exceptional points in higher-dimensional systems.